\documentclass[pre,twocolumn,superscriptaddress,nofootinbib,longbibliography]{revtex4-1}

\usepackage{amsmath}
\usepackage{amsfonts}
\usepackage{amssymb}
\usepackage{array}
\usepackage{dcolumn}
\usepackage{longtable}
\usepackage{hyperref}

\usepackage{float}
\usepackage{bm}
\usepackage{graphicx}
\usepackage{natbib}
\usepackage{color}
\usepackage{Tomspeak}

\begin{document}
\title{Stochastic synchronization induced by noise} 
\date{\today}
\author{Yunxiang Song}
\email{songy1@uchicago.edu}
\author{Thomas A. Witten}
\email{t-witten@uchicago.edu}
\affiliation{Department of Physics and James Franck Institute, University of Chicago, Chicago, Illinois 60637, United States.}

\begin{abstract}

Random perturbations applied in tandem to an ensemble of oscillating objects can synchronize their motion. We study multiple copies of an arbitrary dynamical system in a stable limit cycle, described via a standard phase reduction picture.  The copies differ only in their arbitrary phases $\phi$. Weak, randomly-timed external impulses applied to all the copies can synchronize these phases over time.  Beyond a threshold strength there is no such convergence to a common phase. Instead, the synchronization becomes erratic: successive impulses produce stochastic fluctuations in the phase distribution $q(\phi)$, ranging from near-perfect to near-random synchronization. Here we show that the sampled entropies of these phase distributions themselves form a steady-state ensemble, whose average can be made arbitrarily negative by tuning the impulse strength. A random-walk description of the entropy's evolution accounts for the observed exponential distribution of entropies and for the stochastic synchronization phenomenon.

\end{abstract}
\maketitle
\section{Introduction} \label{sec:Introduction}
Many kinds of signal transmission from cell phone and GPS signaling to magnetic resonance imaging\cite{Callaghan:1993fj}  depend on knowledge of the phase of a remote oscillator.  Because of this, it is valuable to find minimal ways to control such phases.   Remarkably, one can gain useful control by perturbing two oscillators  with random external forcing.  In particular, an ensemble of identical oscillators can be made to synchronize to a common phase by exposing them all to the same random forcing\cite{Pikovskii:1984sh,Pikovsky:1992uu,Pikovsky:1997pj,Teramae:2004qt,Goldobin:2005kn,nakao2007noise}.  Behavior consistent with this mechanism has been reported in cortical neurons\cite{mainen1995reliability} and electronic circuits\cite{Nagai:2009kx}.  Here we examine a useful generalization of this phenomenon in which the resulting synchronization is strong but {\em stochastic}; the ensemble fluctuates between strong and weak synchronization.

We numerically analyzed this stochastic synchronization by repeatedly disturbing the state of the oscillators using a specific common impulsive force at random times, as in Ref \cite{nakao2005synchrony}.  The oscillators are at different positions $\phi$ along their orbits at the moment of the impulse and thus undergo different responses.  Nevertheless, after a transient, each oscillator returns to the limit-cycle orbit at some new position $\psi$ defined below. 
This $\psi$ depends on the orbit position $\phi$ at which the impulse occurred; that is, $\psi$ is a function of $\phi$.  For a given oscillator and a given impulsive force, this ``phase map" $\psi(\phi)$ is a fixed function \footnote{This phase map is denoted $F(\theta)$ in Ref. \cite{nakao2005synchrony}}.  This function suffices to determine the outcome after many impulses. Many synchronization phenomena are amenable to this phase map or ``phase-reduction" analysis. One experimentally important type of forcing is randomly-timed impulses with a common phase map, discussed in Sec \ref{sec:randomTime}.  An initially random set of orbit positions $\phi$ thus repeatedly transforms into a new set.  Synchronization then amounts to progressive bunching of these positions as the iteration proceeds.   By contrast, the stochastic synchronization treated here is more ambiguous.
\begin{figure}[htbp]
\includegraphics[width=\columnwidth]{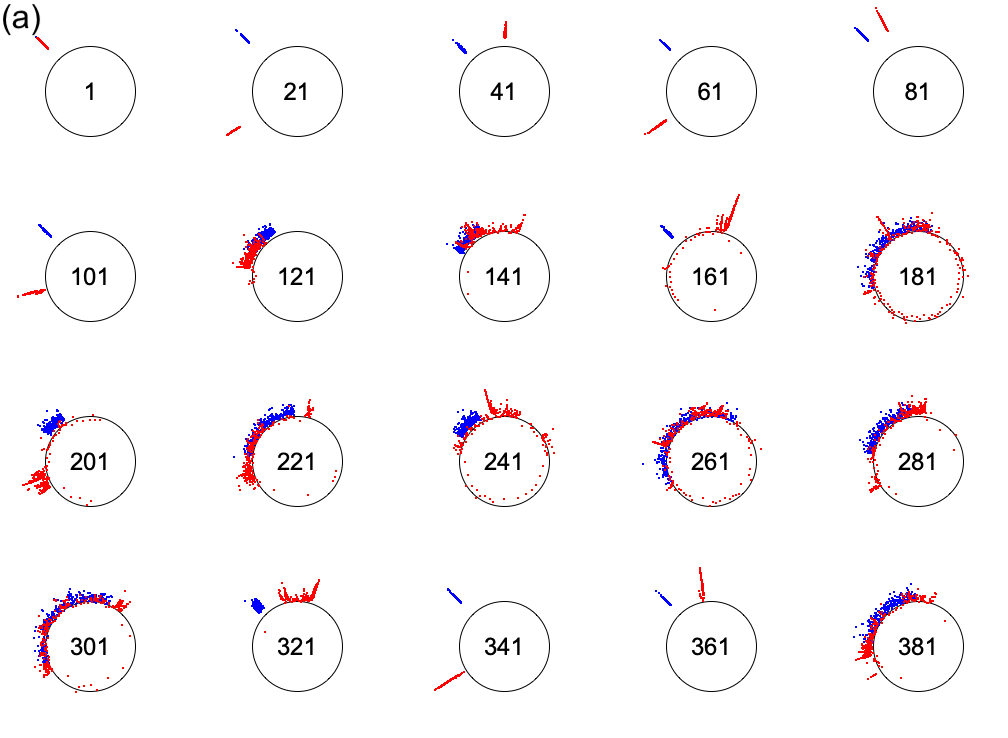}
\bigskip 
\includegraphics[width=\columnwidth]{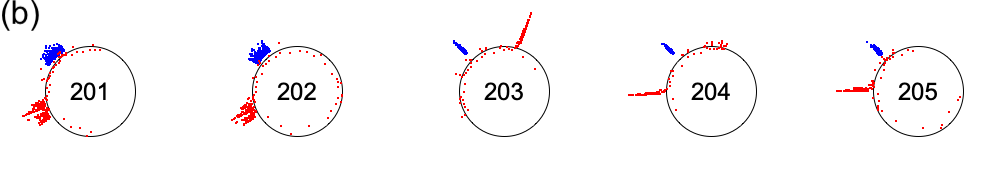}
\caption{\label{fig:redDots} Polar plots of local density of phase angles during iteration using a cubic phase map on the circle as described in Sec. \ref{sec:Simulations}.  The phase map used has the form of Eq. \eqref{eq:cubicPsi} with $A = 9.392$ and $\Phi = 0.445$, for which the Lyapunov exponent $\Lambda \aboutequal +0.141$ (\cf Sec. \ref{sec:randomTime}). The initial ensemble  $\{\phi^1, ..., \phi^{300} \}_0$ consists of 300 positions sampled randomly in an interval of width $0.00025$ at the upper left side of the circle.   For each sample $\alpha$, we denote the separation between $\phi^\alpha$ and its successor around the circle  as $\Delta \phi^\alpha$.  We plot the $\log$ of the sampled density \ie $\log [1/(\Delta \phi^\alpha)~]$ \vs the midpoint position of the pair on the circle as a light-colored (red) dot.  The $\Delta \phi$'s are measured relative to that of uniformly-spaced $\phi$'s. Circle radius corresponds to this uniform spacing.  
 (a): Iterations 1, 21, 41, ..., 381 are shown by rows left to right.  Radial spikes show regions of closely spaced angles, contributing dominantly to the negative entropy $-H$ (Eq. \eqref{eq:sampledH}).  For comparison, dark colored (blue) dots show analogous $\phi$'s sampled from a uniform distribution whose width is chosen to give the same entropy as the phase-mapped $\phi$'s. For definiteness, these distributions are centered at the upper left of the circle, like the initial distribution.
(b):  Detail of iterations 201--205, showing progressive evolution over individual iterations.}
\end{figure}

Figure \ref{fig:redDots} illustrates the distinctive behavior of stochastic synchronization in an ensemble of 300 oscillators subjected to a sequence of 380 randomly-timed impulses, as described above.   Each numbered circle depicts a set of $\phi$'s at some iteration.  Specifically, the light colored (red) dots around each circle depict a sampling of the probability densities $q(\phi)$ inferred from the spacings of adjacent $\phi$'s.  Sharp spikes indicate phase angles having high probability density.  Thus synchronization of the ensemble would appear as a single tall spike at each iteration, showing virtually the same phase angle for all oscillators.  However, in Fig. \ref{fig:redDots} there is no evident convergence to an ordered, synchronized state. Instead, the spikes continually sharpen and dissipate.  Our aim in this paper is to quantify this ongoing stochastic behavior of the ensemble of phases. 

This behavior is distinct from the variety of synchronization phenomena reported in the literature.   It is quite distinct from the familiar synchronization caused by mutual interaction of the oscillators, such as mechanical clocks, fireflies, or applause \cite{pikovsky2003synchronization, winfree1967biological, mirollo1990synchronization}.  Synchronization without such interaction has been studied extensively by phase-reduction methods\cite{Teramae:2004qt,nakao2007noise}.  Noisy perturbations such as weak Gaussian white noise or weak random telegraph noise was shown to lead to complete synchronization.  The breakdown of this synchronization for stronger perturbations has been characterized\cite{Nagai:2005pd,Goldobin:2005kn,pikovsky2003synchronization}.  Our study below uses our methods previously developed to study a colloidal realization of this transition\cite{moths2013full, eaton2016criterion, Witten:2020ls}.

In Sec. \ref{sec:randomTime} to follow we provide a physical context for the random-timed impulses and their statistical description. There we formally define our main analytical tools: the Shannon entropy $H$ and average Lyapunov exponent $\Lambda$. In Sec. \ref{sec:Simulations} we describe the simulations that produced Fig. 1 in terms of $H$ and $\Lambda$, noting that the distribution of $H$ values is a broad exponential.  In Sec. \ref{sec:Explanation} we give an argument to account for this observed behavior of $H$ and predict how this distribution depends on $\Lambda$.  Lastly, in Sec. \ref{sec:Discussion}, we relate stochastic synchronization to similar known phenomena and suggest implications for future work.
\section{random-time impulses and cyclic motion} \label{sec:randomTime}

\subsection{Phase coordinates} \label{sec:coordinates} 
Our study considers physical systems that spontaneously repeat a periodic cycle, such as a dripping faucet, a cooling iron, a beating heart, or a firing neuron\cite{mainen1995reliability}.  Such a system executes a periodic orbit in its dynamical coordinate space in time with some period $T$.  In stable oscillators such as these, a moderate transient disturbance may move its position away from the periodic orbit, but it eventually returns to the orbit.  We define a phase variable $\phi$ by first identifying a convenient starting point on the orbit designated $\phi=0$.  The unperturbed system advances from this point with time $t$ and returns to the $\phi=0$ point in one period $T$.  We then label the orbit point reached at time $t$ by the ``isochronous" phase coordinate $\phi(t) \definedas t/T.$   In a random assembly of such oscillators all times $t$ are equally represented. Therefore the probability distribution of $\phi$, denoted $q(\phi)$  is uniform in $\phi$; \ie $q(\phi) = 1$.  

We may determine the probability $q(\phi)$ by tracking a hypothetical ensemble of many identical copies of the oscillator labeled by $\alpha$ .  If the resulting system is ergodic, this same $q(\phi)$ describes a single oscillator over long times.   Synchronization over time corresponds to a long-time bunching of the various $\phi^\alpha$ towards a common value, which generally varies in time.  The corresponding $q(\phi)$ is sharply peaked. The unperturbed motion simply advances all $\phi$'s around the phase circle at a uniform rate, leaving their relative positions unchanged.  

\subsection{Lyapunov exponent and entropy} \label{sec:Lyapunov}
As noted above, the effect of a momentary disturbance on such an ensemble can be summarized by a ``phase map" $\psi(\phi)$  giving the relative phase long after the disturbance, depending on the phase $\phi$ where the impulse occurred, \cf \cite{nakao2005synchrony}. Specifically. $\psi$ is the final phase relative to that of a copy whose $\phi(t)=0$.
It is convenient to designate the point $\phi = 0$ to be a fixed point where $\psi(\phi) = \phi$.  Then a phase map has  the general appearance of Fig. \ref{fig:Family}, passing through the origin and through the (equivalent) point (1, 1).  Weak disturbances necessarily have phase maps lying close to the $\psi=\phi$ line.  No $\phi$ is changed very much by the impulse, and any bunching effect is weak.  The bunching effect on a small interval of points near a given $\phi$ depends simply on the absolute derivative $|\psi'(\phi)|$ there.  The corresponding interval of $\psi$ is widened by a factor $|\psi'(\phi)|$.  It is useful to define $\lambda(\phi)$ to be the log  of this widening factor.  The average of this $\lambda$ over all $\phi$ is denoted $\Lambda$ and is called the average Lyapunov exponent  \cite{*[]  [{.  Our $\Lambda$ is denoted $\lambda$ in this reference.}]nakao2005synchrony}\cite{eaton2016criterion} 
\begin{equation}\label{eq:Lambda}
\Lambda[\psi(\phi)] \definedas \integral d\phi~ (\lambda(\phi)) = \integral d\phi ~\left[\log  \left |{d\psi \over d\phi} \right | \right ]
\end{equation}
Weak impulses in general have $|\psi'|$ near 1 and thus $\Lambda$ values near zero.   The synchronization properties of a phase map depend strongly on its $\Lambda$ as we 
explain below.

Further iterations of the same type result in the same mapping being applied repeatedly to the ensemble.  The result of such iteration is well documented in textbooks on dynamical systems.  The effects depend on delicate details such as the precise amount of time elapsed between impulses.  However, the effects on disorder are more robust when the amount of time between impulses is random, so that a given oscillator advances an arbitrary fraction of a cycle at the next impulse.  We term this process ``random-shift iteration".  One may express this operation as an alternating iteration of two maps.  The first is the $\psi(\phi)$ described above--- identical for each iteration.  The second is a simple shift map $\tilde\psi(\phi) \definedas \phi + \beta_i$, the same for all $N$ oscillators, but random for each iteration $i$.  Thus the random shifts $\beta_i$ produce a form of noise common to the oscillators. This random shift limit corresponds to Poisson-distributed time between pulses in the limit when their average rate goes to 0, as discussed in Ref. \cite{nakao2005synchrony}. 

To quantify the growth of order or disorder among the phase angles, we adopt the entropy measure that gives the thermodynamic work needed to create the order \cite{Reif:2009ap}.   The entropy $H$ of a probability measure $q(\phi)$ is defined \cite{Shannon_1948}  as 
\begin{equation}\label{eq:Hdef}
H[q(\phi)] \definedas -\integral d \phi ~ q(\phi) ~\log q(\phi) .
\end{equation}

This $H$ has a maximum value of 0 for uniform $q$, \ie  $q(\phi)=1$, with complete disorder.  Conversely, strongly bunched distributions have large, negative $H$.  One may estimate the entropy of an ensemble using a random sampling of $\phi$ values from that ensemble, as detailed below.  

\subsection{Regimes of synchronization} \label{sec:Regimes}
Ref \cite{eaton2016criterion} showed that for random-time iteration of a phase map, the average change of entropy $\expectation{\Delta H}$ in an iteration is bounded above by the Lyapunov exponent $\Lambda$ of the map.  For strongly ordered ensembles with large negative $H$, these authors found that $\expectation{\Delta H}$  in an iteration approaches its upper bound $\Lambda$.  The study of Ref \cite{eaton2016criterion} aimed to test these predictions numerically using realistic phase maps calculated for a specific soft-matter oscillator and a specific type of forcing, ranging from weak to strong.  As the amplitude of the forcing increased from zero, the $\Lambda$ decreased from zero, thus driving $H$ to arbitrarily negative values.  Further increase in the forcing gave a minimum $\Lambda$, where  $H$ decreased most rapidly.  Still further increase in the forcing led to increasing $\Lambda$ and slower decrease in $H$.  Ultimately $\Lambda$ increased to positive values.  For small positive $\Lambda$ initial states with small $H$ initially showed the predicted positive $\expectation {\Delta H}$. But after many iterations the ensemble reached a state of constant average $H$, denoted $\expectation{H}_\infinity$.  The authors noticed that this final $\expectation{H}$ could be made much smaller than the maximum possible $H$ by making $\Lambda$ positive and close to zero.  This is the regime pictured in Fig. 1 that is the focus of the present work.  

\section{Simulations}\label{sec:Simulations}

\subsection{Methods}\label{sec:Methods}

Our simulations aim to explore the transition between small negative $\Lambda$, where $\expectation{H}$ decreases indefinitely, and small positive $\Lambda$, where $\expectation{H}$ is nonzero but seemingly arbitrarily small.  We sought to verify that $\expectation{H}_\infinity$ could be arbitrarily small yet nonzero.  We sought to determine the functional dependence of $\expectation{H}$ on $\Lambda$.  And we sought to explore the generality of this behavior for various map functions.  

Since we aimed to explore general behavior, we used generic smooth phase map functions rather than the experimentally motivated maps of Ref \cite{eaton2016criterion}.  Accordingly, we studied a qualitatively similar cubic function of the form 
\begin{equation} \label{eq:cubicPsi}
\psi(\phi) \definedas  \phi + A \phi (1-\phi)(\Phi - \phi) .  
\end{equation}
Here the amplitude parameter $A$ regulates the distance between $\phi$ and $\psi$, and $\phi = \Phi$ is a fixed point where $\psi(\phi) = \phi$.  By increasing the amplitude $A$ one could increase the value of $\Lambda$.  All $\psi$'s differing by an integer denote the same point on the circle.  

Given this map function, we formally describe the step-by-step procedure for carrying out our numerical simulations. This procedure allows us to judge the dependence on amount of iteration, and its variation with the sequence of random inputs $\beta_i$.
\begin{enumerate}
\item {\bf Ensemble initialization:} An ensemble of $N$ oscillator phases $\{\phi^\alpha\}_0$ where $\alpha = 1, 2, \dots N$ is drawn from a uniform probability density $q_0(\phi)$ supported on the interval $[\phi_0,\phi_0+w)$ on the unit circle of width $w$. The left endpoint $\phi_0$ of this interval is randomly selected.

\item {\bf  Ensemble iteration:} Each phase $\phi^\alpha$ in the ensemble is mapped to an image phase under the phase map $\psi(\phi)$ (Eq. \eqref{eq:cubicPsi}). A random shift $\beta_0$ is applied to all phase in the ensemble to obtain $\phi^\alpha_1 \definedas \psi(\phi_0^\alpha) + \beta_0$, with a corresponding probability density $q_1(\phi)$.

\item {\bf  Entropy estimation:} The entropy $H$ is estimated from the discrete collection of phases $\{\phi^\alpha\}$ \cite{victor2002}. When the number of sampled positions $N$ is large, this estimate can be written
\begin{equation}\label{eq:sampledH}
H\left[ \{\phi^\alpha\} \right] \aboutequal \frac{1}{N} \sum_{\alpha=1}^{N} \log \delta^\alpha + \log\left[2N-2\right] + \gamma ~,
\end{equation} 
where $\delta^\alpha$ is the nearest-neighbor distance for phases $\phi^\alpha$ and $\gamma\aboutequal 0.577$ is the Euler-Mascheroni constant.
\item {\bf  Repeated iteration:} steps 2 and 3 are repeated for $n$ iterations, with a random sequence of shifts $\beta_i$, to generate a single sequence of $\{\phi^\alpha\}_i$ and corresponding entropies $H_i$, for i=1,2, ...,$n$. We typically choose $n$=200.

\item {\bf  Repeated sequence:} step 4 is repeated for $k$ randomly generated sets of $n$ iterations, to obtain $k$ independent trajectories. We typically choose $k$=100.

\item {\bf  Averaging over sequences:} The entropy at each iteration averaged over all $k$ sets of iterations gives a trajectory for the average entropy $\expectation{H}_i$ as a function of iteration $i$.
\end{enumerate}

\subsection{Results} \label{sec:Results}
\begin{figure} [htbp]
\includegraphics[width=3.2in]{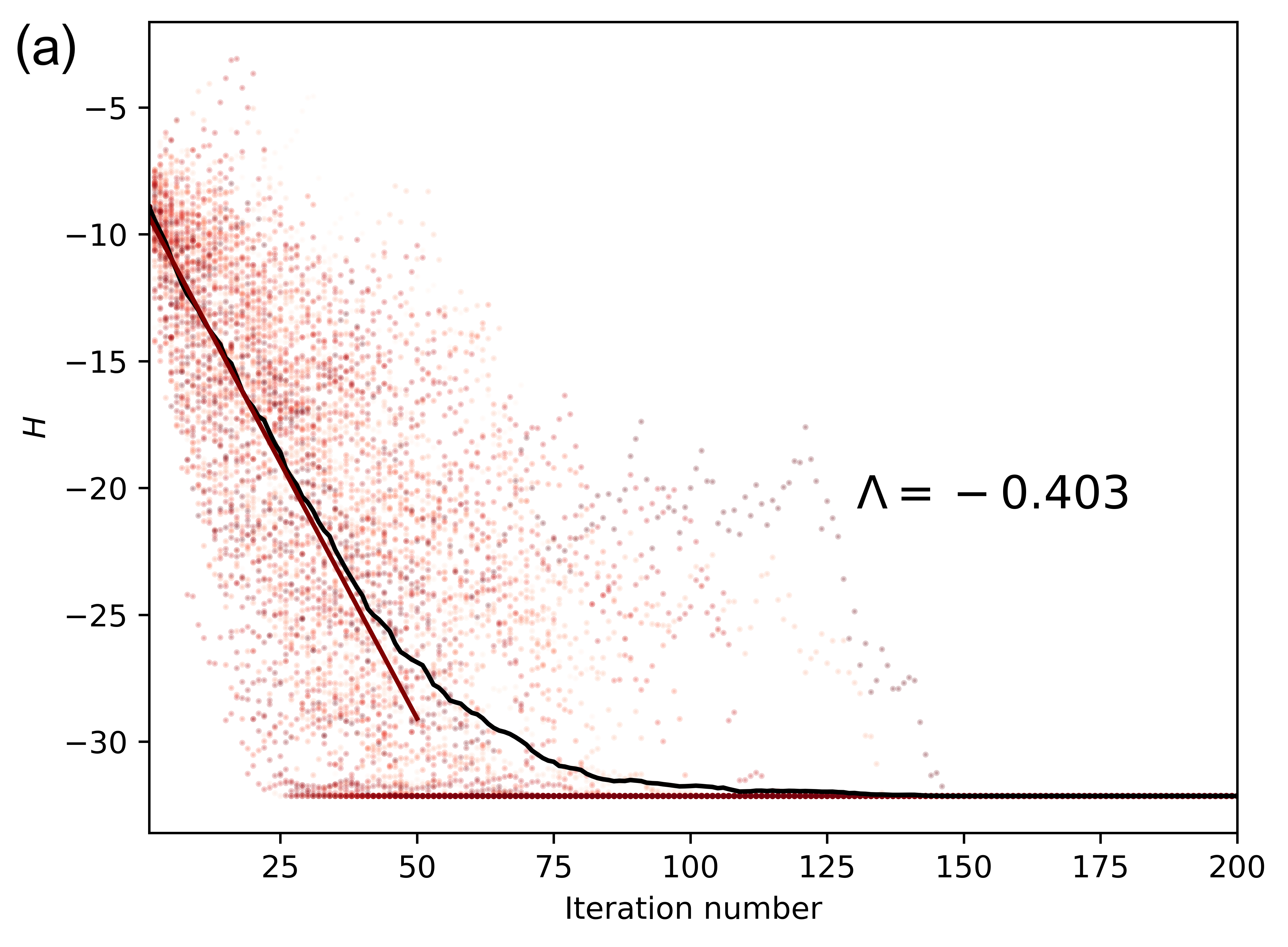}
\includegraphics[width=3.2in]{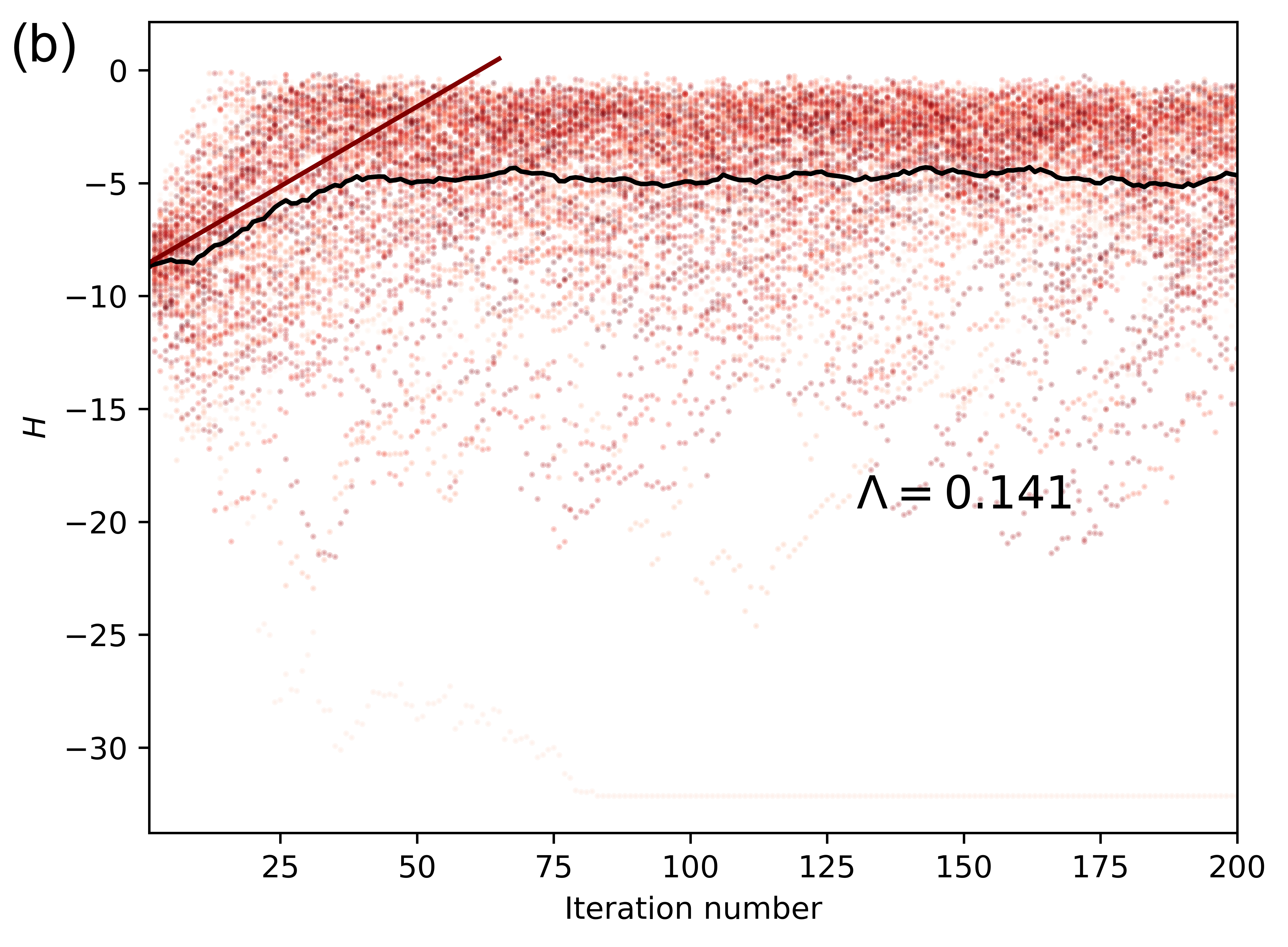}
\includegraphics[width=3.2in]{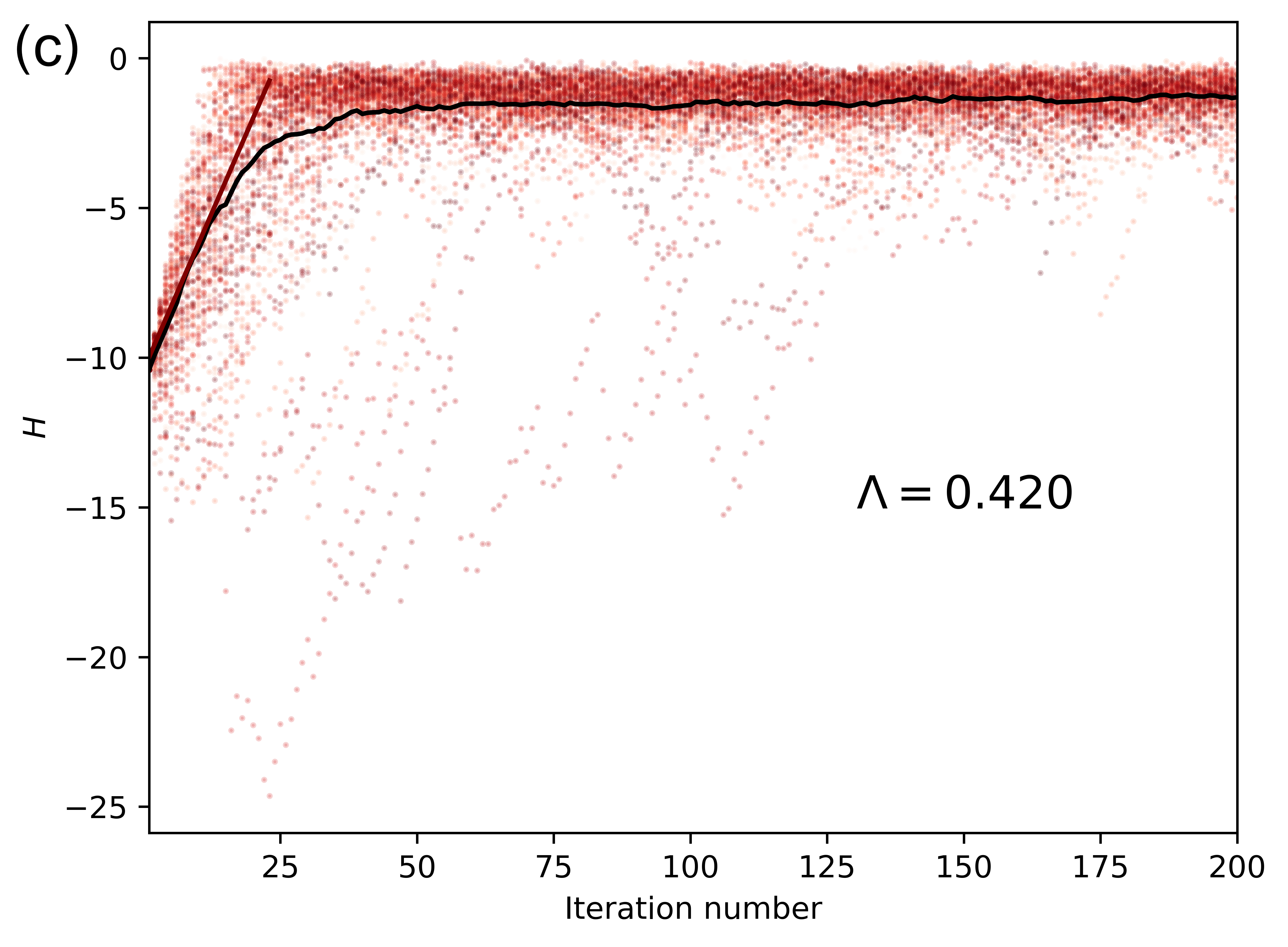}

\caption{\label{fig:HvsiPlots} Entropy $H$ vs iteration number $i$ for randomly-shifted iterated circle maps as described in Sec. \ref{sec:Methods}.  Panels (a)--(c) were made from maps of the form of Eq. \eqref{eq:cubicPsi} with  three different values of Lyapunov exponent $\Lambda$ as indicated.  Each color tracks the entropy change over a single trajectory. Black trajectory shows the entropy at each iteration averaged over all colored trajectories. Straight colored line at left indicates slope $\Lambda$.  Concentration of points for $H \aboutequal -33$ is due to limited machine precision, as described in the text. 
}
\end{figure}

In agreement with \cite{eaton2016criterion} we observe three types of characteristic behavior for iterated phase maps $\psi(\phi)$ with different $\Lambda$. We plot examples of these distinct behaviors in Fig. \ref{fig:HvsiPlots}.

\begin{figure}[htbp]
\includegraphics[width=\columnwidth]{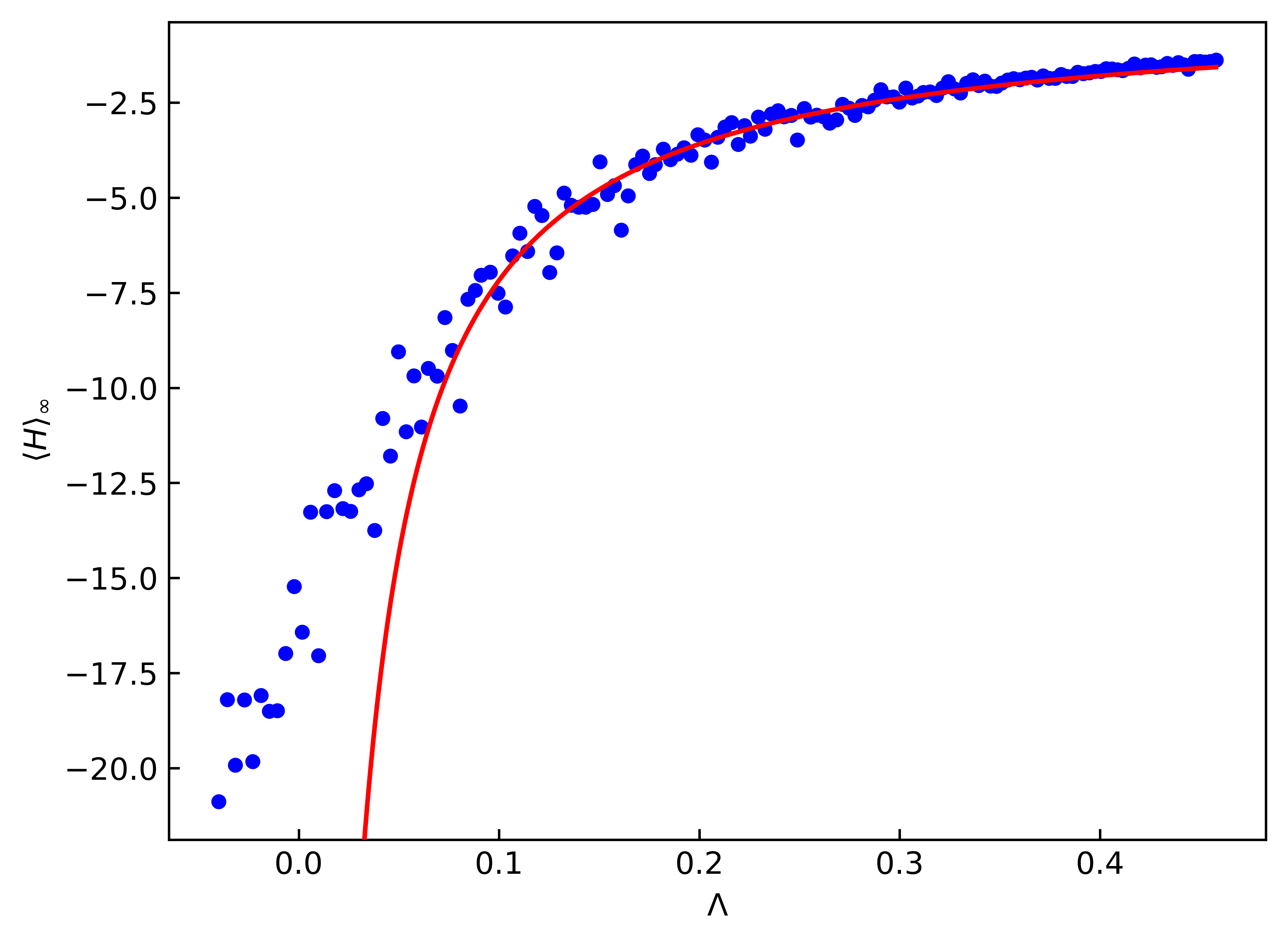}
\caption{\label{fig:AveHvsLambda} Steady-state entropy vs. average Lyapunov exponent.  Dark color (blue) dots plot simulated data for a selection of iterated maps with near continuous range of $\Lambda$. Light color (red) curve is a fit to the form $\expectation{H}_\infinity = a/\Lambda$ anticipated in Sec. \ref{sec:Explanation} with $a= 0.77$.
Data for small $\Lambda$ was subject to distortion in the positive direction owing to our treatment of small separations of the $\{\phi^\alpha\}$ \cf Sec. \ref{sec:Results}.
}
\end{figure}

The first class of behavior is for negative $\Lambda$, as shown in Fig. \ref{fig:HvsiPlots}(a). Here the average entropy decreases to arbitrarily small values over time, so that the system becomes synchronized. The $\expectation H$  decreases at a rate consistent with $\Lambda$, as explained \eg in Ref. \cite{eaton2016criterion}. Here, we note the entropy floor in Figs. \ref{fig:HvsiPlots}a,b just below $H=-30$ instead of an indefinite decay towards $-\infinity$. Such a floor is an artifact of our simulation implementation. When two neighboring phase angles are indistinguishable up to machine precision, we manually separate them by 1$\times 10^{-15}$ to maintain finite values in the the sum of Eq. \eqref{eq:sampledH}.  The floor is reached when all phase angles in the ensemble are indistinguishable.  These numerical limitations produced significant distortions in our measured $\expectation{H}_\infinity$ for $\Lambda \lessthanorabout 0.14$ (\cf Fig.\ref{fig:AveHvsLambda}).

The second class of behavior is for large positive $\Lambda$, as shown in Fig. \ref{fig:HvsiPlots}(c). In this case, the average entropy approaches the maximally disordered state with $\expectation H \aboutequal 0$. Such behavior is also well understood \cite{eaton2016criterion,Witten:2020ls}. 

The third class of behavior exists in a narrow regime between the previous two classes, where $\Lambda$ is small, but still positive, so that there is an initial upwards slope in $\expectation{H}$, as shown in Fig. \ref{fig:HvsiPlots}(b). The average change in entropy initially follows the upper bound $\Lambda$ but eventually reaches a steady state. The small asymptotic value $\aboutequal -5$ indicates states that are on average much more ordered than a completely random set of phases.  
 
\begin{figure}[htbp]
\includegraphics[width=\columnwidth]{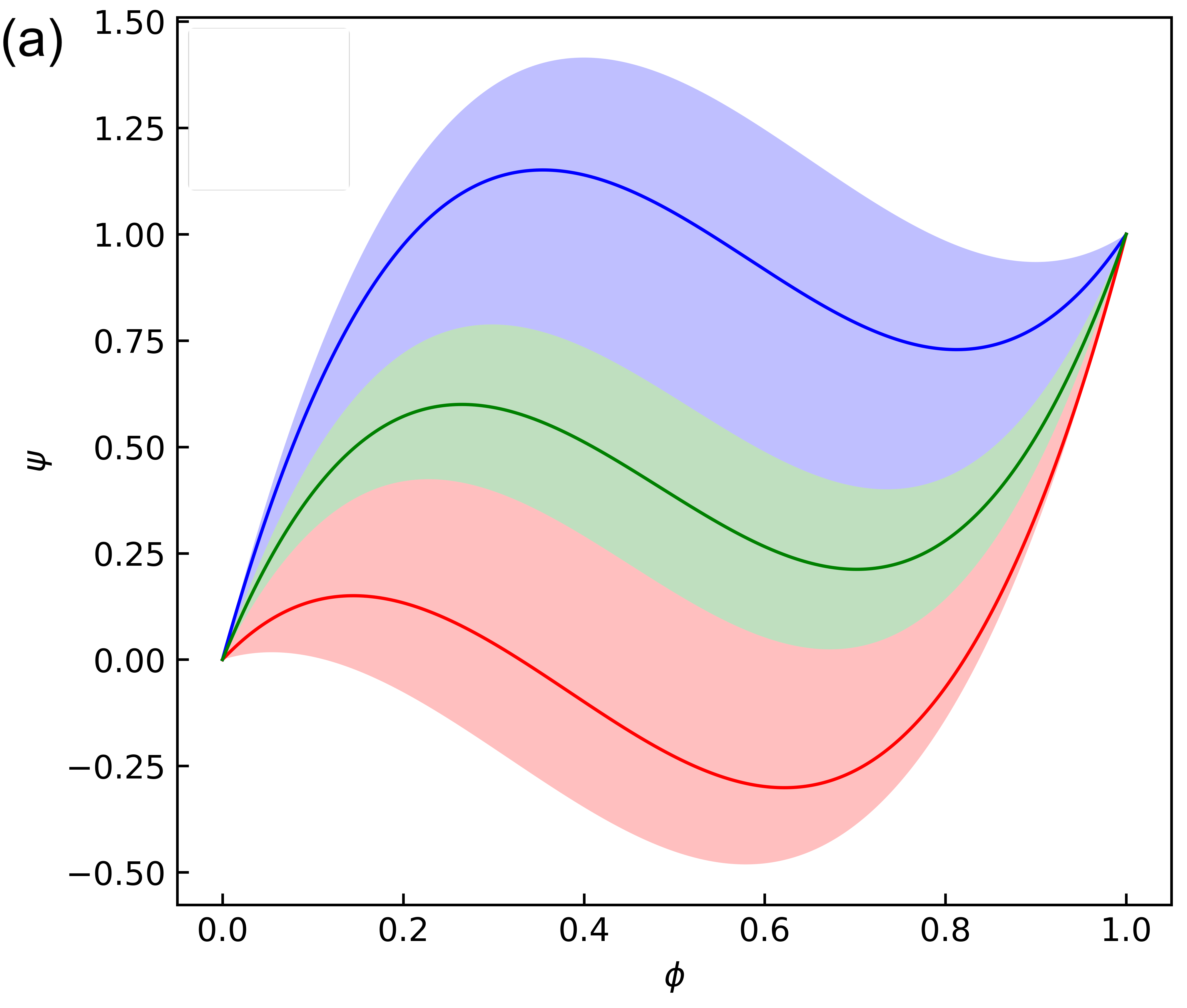}
\caption{\label{fig:Family} Phase maps $\psi(\phi)$ of the form of Eq. \eqref{eq:cubicPsi}, used to study the effect of varying the phase map with fixed $\Lambda \aboutequal 0.141$.  Fixing $\Lambda$ allows one free parameter within the class of cubic maps.  The $\expectation{H}_\infinity$ values obtained from simulating 30 maps spanning this range was the same within the statistical uncertainty, \viz  $\expectation{H}_\infinity = -5.5 \plusorminus 0.2$ .}
\end{figure}

\begin{figure} [htbp]
\includegraphics[width=\columnwidth]{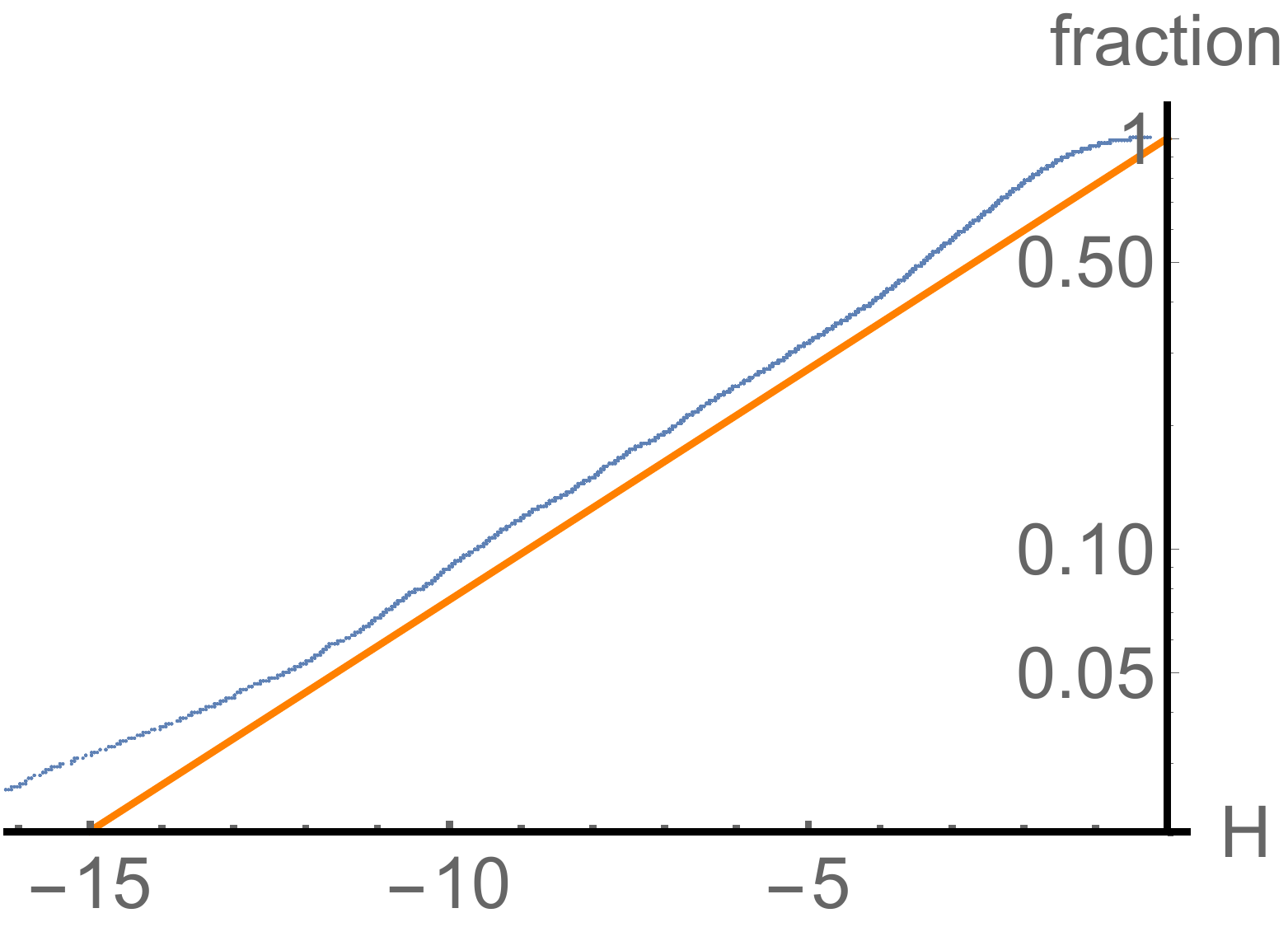}
\caption{\label{fig:PofHfig} Cumulative distribution of $H$ as obtained in Sec. III under conditions of Fig.\ref{fig:HvsiPlots}b, using a map with Lyapunov coefficient $ \Lambda = 0.14$, shown as a semilogarithmic plot.  The $P(H)$ defined in the text is the derivative of the cumulative distribution plotted.  Dark line shows for each entropy $H$ the fraction of samples with entropy smaller than $H$.  Mid section of this curve is approximately  a straight line indicating exponential falloff.  Light line is the exponential distribution with a scale height $H_0$ matching this section: $H_0 = 3.85$ .}
\end{figure}

\begin{figure}[htbp]
\includegraphics[width=\columnwidth]{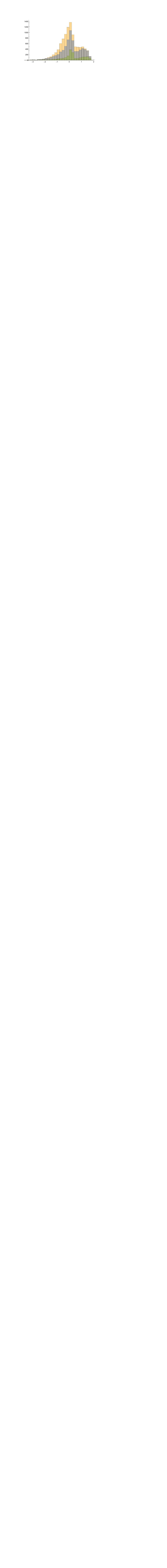}
\caption{\label{fig:hHistograms} Histograms of the distribution of incremental changes of $H$, $p(h)$ for the map with $\Lambda = 0.14$.  Light-colored histogram includes the full sample.  Middle color and foreground color show the half of the sample with smallest $H$ and the 20 percent of the sample with the smallest $H$.  The variances of all the samples were approximately unity.  The mean for the whole sample was approximately 0, as expected for a steady-state process.  For the foreground sample, the mean $\expectation{h}\aboutequal 0.16$, is equal to $\Lambda$ within sampling uncertainty.  The two sharp dropoffs near $h = 0.4$ and $h = 1.8$ correspond to extremal slopes in $\psi(\phi)$.  
}
\end{figure}

We made further statistical measurements of the entropy in this steady-state regime of small average entropy.  First, we verified that $\expectation{H}_\infinity$ indeed extrapolates to $-\infinity$ as $\Lambda \goesto 0$ in the undistorted range $\Lambda > 0.14$ (Fig. \ref{fig:AveHvsLambda}).  The limiting behavior was consistent with $\expectation{H}_\infinity \goesas 1/\Lambda$.  Changing the map while keeping $\Lambda$ fixed had little effect on $\expectation{H}_\infinity$, as shown in Fig. \ref{fig:Family}. Next, we measured the distribution of entropy values $P(H)$, shown in Fig. \ref{fig:PofHfig}.  The measurements strongly indicate an exponential falloff with $|H|$.  Finally, we measured the statistics of the incremental change $h_i \definedas H_{i+1} - H_i$, as shown in Fig. \ref{fig:hHistograms}.  For the most negative range of $H$, the mean $\expectation{h}$ of this distribution approached $\Lambda$, as anticipated \cite{eaton2016criterion}, while the variance was of order unity.   The $h$ values fell off sharply on both sides of the mean, with no apparent long tails.


\section{Explanation of $H$ Distribution in steady state}\label{sec:Explanation}

Here we show that the regime of stochastic fluctuations of the phases with indefinitely small entropy is to be expected for general random-shift phase maps with small positive $\Lambda$.  We  present a scheme that accounts for this behavior.  It accounts for the observed broad  distribution of $H$ values in the steady state.  It also predicts that the average  $\expectation{H}_\infinity$ varies as $1/\Lambda$ for $\Lambda\goesto 0$.  

 Both the asymptotic scaling of $\expectation H$ for small Lyapunov exponent $\Lambda$ and the functional form of the distribution $P(H)$ can be understood in stochastic terms:  the fluctuations of $H_i$ may be viewed as a random walk process.  In this regime, the typical $H$ values and their average are indefinitely smaller than the values observed for large $\Lambda$.  Our description exploits the following features of this small-$\Lambda$ regime: 
\medskip
\begin{enumerate}
	\item The incremental changes in $H$ in one iteration $i$, $h_i \definedas H_{i+1} - H_i$, vary significantly, but the range of this variation is limited and is little dependent on $H_i$ when $H_i \muchlessthan 0$.
	
	\item The mean value of the increment $h$ approaches $\Lambda$ as $\Lambda\goesto 0$, as shown in Ref \cite{eaton2016criterion}.
	
	\item $H$ can never exceed a maximum, namely its value for a uniform probability distribution of phase angles.  With our conventions this maximum is 0.
\end{enumerate}
 
These conditions suggest a simple diffusion-drift mechanism \cite{Einstein:1905fx, randomWalksText} for the strong, steady-state entropy fluctuations that we observe.  We picture the changes of $H$ as a slightly biased random walk, whose steps $h_i$ are drawn independently from a  distribution $p(h)$.  This $p(h)$ is taken to be independent of $H$, as motivated by feature 1) above.  The mean of this distribution $\expectation h$ approaches $\Lambda$, as dictated by feature 2.  Finally, any step that leads to $H > 0$ is withdrawn, as dictated by feature 3.   In the absence of the drift, the entropy undergoes unbiassed excursions and ultimately reaches indefinitely negative values.  However, with small positive $\expectation h$, these excursions are opposed by the drift.  Then $H$ can no longer decrease indefinitely, and a steady state is reached in which  the drift is balanced by the random steps.  This confinement is weaker and $\expectation H$ is more negative as $\expectation h \goesto 0$.  Thus in the small-positive-$\Lambda$ regime of interest, $\Lambda^2 \muchlessthan \expectation{h^2} \muchlessthan \expectation{H}_\infinity^2$.

The behavior of such diffusion-drift processes is well known \cite{randomWalksText} .  In particular, in steady state, the probability distribution $P(H)$ obeys 
\begin{equation} \label{eq:DiffusionDrift}
{d^2 P\over d~H^2} = {2 \expectation h\over \expectation{h^2}} ~ {dP\over d H} ~~,
\end{equation}
so that
\begin{equation}\label{eq:PofHmodel}
P(H) = C~ \exp{(H/H_0)}~~ ,
\end{equation}
where $C$ is a normalizing constant and the scale height $H_0$ is given by 
\begin{equation} \label{eq:ScaleHeight}
H_0 = \expectation{h^2}/(2\expectation h).
\end{equation}

This predicted behavior is consistent with the simulations described above.  Fig. 5 shows the observed distribution of $H$ values for $\Lambda = 0.14$ from Sec. III, incorporating 300 runs of 200 time steps apiece.  Time steps before step 100 were discarded to avoid effects of the initial transient; the remaining sample contained $10^4$ measurements. The plotted distribution is the fraction of samples with entropy more negative than $H$.  The picture shows a one-decade range consistent with an exponential distribution as anticipated in Eq. \ref{eq:PofHmodel}, implying $H_0 \aboutequal 3.85$ .

For a more explicit test of the explanation, we measured the distribution of step changes $h$ observed in the simulation. In general this distribution depends on $H$. Since good prediction is only expected for $H \muchlessthan -\sqrt{\expectation{h^2}}$, we measured incremental $h$ values only for the smallest 20 percent of the $H$ values, as shown in Fig. 6.  Here we obtained $\expectation{h}\aboutequal 0.16$. This value violates the upper bound \cite{eaton2016criterion} on the true average relative to the Lyapunov coefficient $\Lambda$ of the map defined in Eq. \eqref{eq:Lambda}, \viz,  $\expectation{h} < \Lambda = 0.14$. We attribute this discrepancy to statistical uncertainty in the simulation \footnote{
According to the diffusion-drift picture, we expect an uncertainty in $\expectation{h}$ of order $\expectation{h^2}/\sqrt{N}$, where $N (=2000)$ is the number of samples.  Thus uncertainty is roughly 0.02, comparable to the discrepancy between 0.16 and 0.14.  
}.
Thus, we regard the measured $\expectation{h}$ as a crude consistency check.  The variance $\expectation{h^2}$ was approximately $ 1.0$.  These figures give an expected $H_0$ using Eq. \ref{eq:ScaleHeight}: $H_0 = 3.1$ using the measured $\expectation{h}$, or $H_0 
= 3.5$ using $\expectation h = \Lambda$. 
This is roughly consistent with the observed $H_0 = 3.85$ shown in Fig 5, thus lending support to the diffusion-drift picture.

The diffusion-drift picture also accounts for the dependence of $\expectation H$ on $\Lambda$ reported in Sec. \ref{sec:Simulations}.  For asymptotically small $\Lambda$,  we may replace $\expectation h$ by $\Lambda$. The diffusion-drift picture then gives an exponential distribution $P(H)$ with scale height $H_0 \goesto \expectation{h^2}/(2\Lambda)$ and a mean value $\expectation H \goesto -H_0$, which varies as $\Lambda^{-1}$, in agreement with Fig. \ref{fig:AveHvsLambda}. 

Our actual $\Lambda$ values did not reach this asymptotic regime.  Even our smallest $\Lambda$ simulation had a $P(H)$ that departed significantly from an exponential as seen in Fig. \ref{fig:PofHfig}.  It had a mean $\expectation H = -4.8$ that was somewhat smaller than  $-H_0 ~ (=-3.85)$.  The predicted $\expectation{H} \goesto \expectation{h^2} /(2\Lambda)$ is qualitatively consistent with the measurements of Fig \ref{fig:AveHvsLambda}. We verified that $\expectation{h^2} \aboutequal 1$ for large negative $H$ over the range of $\Lambda$ studied.  Thus the asymptotic formula predicts $\expectation H$ of the form $a/\Lambda$, in agreement with the fitted curve in Fig. \ref{fig:AveHvsLambda}.  However, the predicted value of the coefficient $a$ is 0.5---only 64 percent of the fitted value of $a$.

Overall, the asymptotic diffusion-drift picture appears to account adequately for the limited simulations reported above.  In the next section we discuss the generality of this picture.

\section{Discussion}\label{sec:Discussion}

In the foregoing we have demonstrated a novel mechanism for synchronizing an ensemble of oscillators with random phases using a robust, physically applicable forcing protocol.  The procedure can produce controlled disorder of nonzero magnitude as well as indefinitely small disorder.  In this discussion section we justify why our low-entropy states can be viewed as a form of synchronization.  We argue that our system achieves a statistically well-behaved though unusual steady state, and cite a similar established example.  Finally we review the practical limitations of the method, and survey possible generalizations.  

\subsection {Entropy and synchronization}\label{sec:EntropySynchronization} 
Synchronization is conventionally quantified by the uncertainty spread in between the phase angles of two oscillators.  However a single narrow interval of uncertainty is not necessary  in order to have useful knowledge about a phase.  If the probability is divided into two narrow intervals, the phase information does not degrade greatly simply by separating the two intervals by a large spacing.  Indeed, the number of trials needed to determine the phase angle to a given tolerance is equal for the single interval and the split interval regardless of the size of the split.  The difference in entropy of two distributions is precisely the relative number of trials needed on average for the two.  Thus for any given distribution $q(\phi)$ one may identify a uniform distribution that is equivalent in number of trials needed to determine the phase.  This is the uniform distribution whose entropy is equal to that of $q(\phi)$.  For many purposes of information transmission, this entropy measure is what determines the transmitting capacity \cite{Shannon_1948}.  

\subsection{Statistical regularity}  \label{sec:Statistical}
The erratic variability of the synchronization and of the entropy in our system raises the question whether the process is even statistically well-defined.  Several
tests gave us positive reassurance on this point. The behavior of the simulated entropy fluctuations appears consistent with a stationary distribution $P(H)$.  On the one hand, the measured average entropy showed a variance that decreases inversely with number of iterations or trials, as expected for averaging from a fixed ensemble.
 
On the other hand the distribution of $\lambda(\phi)$ ($\definedas \log |\psi'(\phi)|$) values for generic functions $\psi(\phi)$ is also consistent with a well-defined  $P(H)$ distribution.  When $H_i$ is small, this $\lambda(\phi)$ determines the distribution of $H_{i+1}$ values in one iteration starting from a given distribution $q_i(\phi)$. The particular value of $H_{i+1}$ obtained depends on the randomly chosen shift amount $\beta_i$ (Sec. \ref{sec:Lyapunov}).  This dependence can be understood in the limit where $q_i(\phi)$ has support confined to a few narrow intervals on the circle.  Then the $H_{i+1}(\beta_i)$ for all $\beta_i$ can be readily inferred from the properties of $\lambda(\phi)$, as shown in the Appendix.  In the opposite regime of uniform $q_i$, there is no $\beta$ dependence of $H_{i+1}$ and thus no randomness in the distribution of $h$.  Thus the assumption of well-behaved randomness of $h$ is confirmed for the highest as well as the lowest initial entropy $H_i$.  

\subsection{Parallels} \label{sec:parallels} 

Stochastic synchronization extends the notion of ``noisy on-off intermittency" or ``bursting behavior" reported in the noise-induced synchronization literature \cite{Teramae:2004qt, nakao2005synchrony, nakao2007noise}.  This behavior was seen in oscillators subject to two types of random perturbation: (a) common impulses as described above with {\em negative} Lyapunov exponent and (b) uncorrelated Langevin noise.  There the intermittency arose from the competition between the disruptive influence of the noise and the synchronizing effect caused by the local Lyapunov exponent $\lambda(\phi)$ over many iterations.  Our work shows that this intermittent steady state extends to positive $\Lambda$, where the disordering effect of $\Lambda$ is opposed by the ordering effect of any phase map on a near-uniform distribution $q(\phi)$.  

More broadly, bursting intermittency is encountered in dynamical systems near a transition to chaos.  Properties of this intermittency such as scaling of correlations and size of the chaotic region have been explained using properties of the incipient chaotic state \cite{Heagy:1994yx,Venkataramani:1996ty,Yu:1990iu}.  The present work doesn't require proximity to a chaotic state and uses only the phase map associated with some perturbation.  Nevertheless, the nature of our disorder appears similar to that seen in the prior works.  It seems likely that the degree of disorder in these near-chaotic systems quantitatively resembles that shown in the present system, that operates well within the boundaries of a stable limit cycle.

Stochastic states analogous to ours are well known in disordered wave systems.  There, as in our system, 
 configurational variables analogous to $\phi$ lack well-defined probabilities that reveal the statistical behavior of the system.  One case of similar behavior is the transmission of a wave such as light through a stack of $n$ different transmitting layers such as sheets of glass of different 
 thicknesses \cite{StoneJoannopoulos1981, Chabanov:2000dq}. 
 If the sheets are all much thicker than a wavelength, the complex reflection and transmission amplitudes vary widely from sheet to sheet.  As a result, the overall transmission varies erratically with wavelength, with no single characteristic distribution.  The distribution of transmission coefficients is ill-conditioned and only the logarithm of the transmission coefficient has well-conditioned statistics.  In our system the probability distribution $q(\phi)$ is ill-conditioned and the entropy does not reach a definite value for large systems.  Instead the entropy converges to a steady-state distribution with a well-determined average.  

\subsection{Limitations}\label{sec:Limitations} 
In investigating this class of stochastic dynamics we encountered limitations in exploring the asymptotic regime of strong but incomplete ordering.  Numerically, we were unable to reliably explore maps with positive Lyapunov exponents $\Lambda$ much smaller than 0.14 and corresponding $\expectation H \lessthanorabout -5$.  The difficulty is already apparent in Figs. \ref{fig:HvsiPlots}a,b.  There, as noted in Sec. \ref{sec:Results} one sees several trajectories that descend to the bottom of the figure and remain there.   These trajectories have $\phi^\alpha$ values separated by less than the machine precision of the calculations. Such pairs of $\phi$'s cannot give valid contributions to Eq. \eqref{eq:sampledH}. These trajectories necessarily increase in number as the iterations proceed, eventually compromising the measurement of entropy.  Thus the number of iterations attainable in practice is limited.  This in turn limits the $\Lambda$ values that can be explored, since small $\Lambda$ entails slow relaxation to the steady state.  These numerical limitations can be mitigated by increasing the numerical precision of the simulation. In this way one could improve our crude validation of the mechanism described in Sec. \ref{sec:Explanation} . This improvement would be very desirable.

Similar limitations would be expected in experiments like the soft-matter experiments treated in Ref. \cite{eaton2016criterion}.  First, the final state of any dynamical oscillator is subject to random noise as well as to the imposed random-time phase maps.  This noise requires the forcing to be sufficiently strong to induce rapid synchronization.  Any synchronization is thus limited by the random noise \cite{nakao2007noise}.  A second limitation comes from the inevitable variability of colloids or other oscillators in the ensemble.  This variability means each element $\alpha$ of the ensemble has a distinct phase map  $\psi_\alpha(\phi)$.

Such limitations mean that clear-cut experimental attainment of the asymptotic predictions above is unlikely.  Nevertheless, this limit could be useful for devising well-behaved statistics to describe the stochastic synchronization regime, as was the case with disordered quantum wires in Ref. \cite{StoneJoannopoulos1981}.  

\subsection{Generalizations} \label{sec:Generalizations}
The stochastic dynamics treated here was chosen for its simplicity. It treats only a very restricted type of disruption of the limit cycle. Still, the anomalous ordering behavior demonstrated here should occur more generally.  Many of the known noise-induced synchronization phenomena\cite{Teramae:2004qt, nakao2005synchrony, yoshida2006noise, nakao2007noise, hata2010synchronization}  have been demonstrated for broad classes of noise, including continuous random driving.  These phenomena are regulated by an average Lyapunov exponent $\Lambda$ analogous to ours.  Generally synchronization was reported only when this $\Lambda$ was negative.  Our results suggest that a form of stochastic synchronization can occur even when $\Lambda$ is positive.  

For some incremental generalizations of our simple forcing, stochastic synchronization like that shown here appears likely.  For example, if the impulsive forces are not restricted to be identical, but instead vary in strength, the ordering should persist.  Such variable forcing would lead to a range of phase maps $\psi_i(\phi)$ for different iterations $i$.  Still if these maps all had $\Lambda$ values bounded above by a small, positive number, one would expect bounds on the entropy similar to those shown here.  Likewise, we restricted the impulses to be widely separated in time so that the oscillators all relax to their periodic cycles before the next impulse.  This requirement simplifies our analysis but does not appear essential to the phenomenon of stochastic ordering \cite{nakao2005synchrony}.  

One general virtue of noise-induced synchronization is that one may use it to induce synchronization without specific knowledge about the limit cycle being synchronized, such as the phase map function or the cycle time.  We note that this virtue is preserved for the impulsive noise of the current study.  

\section{Conclusion} \label{sec:Conclusion}

As noted in the Introduction, synchronization of a remote oscillator enables transmission of information.  Now, our demonstration of low-entropy oscillator states is far from a demonstration of effective information transmission.  Yet this stochastic counterpart of synchronization shows potential as a generalized means of transmission, potentially useable by technology or living systems.  Further, this model of intermittency may offer a tool to address open questions in wave localization and strong turbulence.


\section*{Acknowledgments} 

We thank Ivar Martin, Peter Littlewood, Kyle Kawagoe, and Alex Edelman for insightful discussions.  Jonah Eaton and Martin Falk suggested valuable improvements in the manuscript.   Y.S. acknowledges support from the University of Chicago Jeff Metcalf Internship Program.

\section*{Appendix: Limits on the range of $H_{i+1}$ values}\label{sec:appendix}

Here we address the claim made in Sec. \ref{sec:Statistical} that the change in entropy under a single iteration of our random-time phase mapping varies within a limited range.  We consider the regime where the initial distribution $q_i(\phi)$ is narrowly defined, as is typical when the initial entropy $H_i$ is small. In general the new probability distribution $q_{i+1}(\hat\phi)$ can be expressed \cite{eaton2016criterion} as
\begin{equation}\label{eq:qitoqip1}
q_{i+1}(\hat\phi) = \integral d\phi ~q_i(\phi) ~\delta\left (\hat\phi - \psi(\phi+\beta_i) \right ) .
\end{equation}
Then $H_{i+1}$ is obtained from $q_{i+1}(\hat\phi)$ by applying Eq. \eqref{eq:Hdef}.
We note that when $\psi(\phi)$ is non-monotonic, several values of $\phi$ may contribute to a given $\hat \phi$.  

As announced in the main text, we first consider the range of $H_{i+1}$ values when $q_i$ is  confined to an arbitrarily small support.  For such ``well-ordered" $q_i(\phi)$ there is generally only one $\phi$ contributing to each $\hat\phi$ in Eq. \eqref{eq:qitoqip1}. Then Eq. \eqref{eq:qitoqip1} becomes 
\begin{eqnarray}\label{eq:qip1}
q_{i+1}(\hat\phi) = q_i(\phi)&~ [\integral d\phi ~\delta(~\hat\phi - \psi(\phi + \beta_i) ~)] \nonumber\\
= q_i(\phi)& ~|\psi'(\phi + \beta_i)|^{-1} .
\end{eqnarray}

For such $q_{i+1}$, $H_{i+1}(\beta_i)$ becomes a convolution of $\lambda(\phi + \beta)$ with $q_i(\phi)$.  To show this\cite{eaton2016criterion}, we use the definition from Eq. \eqref{eq:Hdef} with Eq. \eqref{eq:qip1} and the fact that $d\hat \phi ~q_{i+1}(\hat \phi) = d \phi ~q_i(\phi)$ to    write $H_{i+1}$ as 
\begin{equation}
H_{i+	1}(\beta_i) = -\integral d\phi ~ q_i(\phi)~ \log \left (~ |\psi'(\phi + \beta_i)|^{-1}~ q_i(\phi)~\right ) .
\end{equation}
After decomposing the $\log$ and noting that $\log(|\psi'(\phi + \beta_i)|) \definedas \lambda(\phi + \beta_i)$, this gives
\begin{equation}\label{eq:Hip1}
H_{i+1}(\beta_i) = \integral d\phi ~ q_i(\phi) ~\lambda(\phi + \beta_i) - \integral d\phi ~ q_i(\phi) ~ \log q_i(\phi) . 
\end{equation}
Here the second term (simply $H_i$) is independent of $\beta_i$ while the first term is the convolution claimed.

We now argue that the range of $H_{i+1}(\beta_i)$ values is strongly limited.  This range is greatest when $q_i(\phi)$ is simply a delta function.  Then, ignoring the constant second term in  Eq. \eqref{eq:Hip1}, the distribution of $H_{i+1}$ values is simply the distribution of $\lambda$ values.  This distribution may be readily evaluated for any given phase map $\psi(\phi)$.  For a smooth map function like those encountered in \cite{eaton2016criterion} and those considered here, there is a maximum $\lambda$ at the point of maximum absolute slope.  There is no minimum value since there are  extrema where $\phi$ has arbitrarily small $|\psi'|$.  We consider generic $\psi(\phi)$ with no points where $\psi'$ and $\psi''$ simultaneously vanish, so that all extrema of $\psi(\phi)$ are quadratic.  
Near any extremum $\phi^*$, $|\psi'| \goesas |\phi-\phi^*|$, so that $\lambda(\phi) \goesas \log|\phi-\phi^*|$.  Then the $p(\lambda)$ falls off exponentially for large negative $\lambda$. 
 Thus for well-ordered $q_i(\phi)$ the range of $H_{i+1}$ is well-confined, the central limit theorem\cite{randomWalksText} applies, our diffusion hypothesis of Sec. \ref{sec:Explanation} is well justified and a well-defined $P(H)$ is reached.  

		The range of $H_{i+1}(\beta_i)$ is also narrow in the opposite limit of uniform $q_i(\phi)$.  Here the shift $\beta_i$ has no effect on $q_i(\phi)$ and thus $H_{i+1}(\beta_i)$ is independent of $\beta_i$; The spread of $H_{i+1}$ values for a given $H_i$ goes to zero.  While these arguments cover only the limiting cases and fall short of a proof, they lend plausibility to our numerical finding that the distribution of entropy increments $h_i$ is statistically well-behaved. 
\vfill\eject
\bibliography{AsymptoticH}

\end{document}